\documentclass[aps,prl,twocolumn,groupedaddress,longbibliography]{revtex4-1}

\usepackage{graphicx}

\begin{document}
	

\title{Slow decay processes of electrostatically trapped Rydberg NO molecules}


\author{A. Deller}
\altaffiliation{Present address: Max-Planck-Institut f\"ur Plasmaphysik, Boltzmannstra{\ss}e 2, 85748 Garching bei M\"unchen, Germany.}
\author{M. H. Rayment}
\author{S. D. Hogan}

\affiliation{Department of Physics and Astronomy, University College London, Gower Street, London WC1E 6BT, United Kingdom}
	

\date{\today}
	
\begin{abstract}
Nitric oxide (NO) molecules initially traveling at 795~m/s in pulsed supersonic beams have been photoexcited to long-lived hydrogenic Rydberg-Stark states, decelerated and electrostatically trapped in a cryogenically cooled, chip-based transmission-line Rydberg-Stark decelerator. The decelerated and trapped molecules were detected \emph{in situ} by pulsed electric field ionization. The operation of the decelerator was validated by comparison of the experimental data with the results of numerical calculations of particle trajectories. Studies of the decay of the trapped molecules on timescales up to 1~ms provide new insights into the lifetimes of, and effects of blackbody radiation on, Rydberg states of NO. 
\end{abstract}
	
	
\maketitle
	
Molecules in high Rydberg states play important roles in tests of fundamental physics through measurements of ionization and dissociation energies~\cite{holsch19a}. Their decay dynamics affect recombination in atmospheric and astrophysical plasmas~\cite{peverall01a,guberman07a}, while in the laboratory they have been exploited in studies of ultracold plasmas~\cite{morrison08a,sadeghi14a}. Samples in these states are also of interest for low-energy inelastic scattering including, e.g., studies of long-range dipole-dipole interactions that lead to resonant energy transfer~\cite{smith78a,safinya81a,gawlas20a}, low-energy electron scattering that can allow the stabilization of weakly bound long-range Rydberg molecules~\cite{greene00a,bendkowsky09a} and result in electron transfer~\cite{kelley17a,engel19a}, and short-range ion-molecule reactions~\cite{allmendinger16a}. In these areas, advances are expected if the molecules are prepared in slowly moving, velocity-controlled beams or in traps. This would enhance resolution in spectroscopy and scattering experiments and facilitate direct measurements of lifetimes and decay dynamics over previously inaccessible timescales.

Experimental methods -- including, e.g., buffer gas cooling~\cite{doyle95a}, multistage Stark~\cite{bethlem99a,wang13a} and Zeeman~\cite{vanhaecke07a,narevicius08a} deceleration, and laser cooling~\cite{shuman09a,shuman10a} developed for the direct preparation of cold ground-state molecules offer opportunities for studies involving high Rydberg states~\cite{jansen18a}. However, the most widely applicable approach to preparing cold, trapped Rydberg molecules involves exploiting the forces that can be exerted on them using inhomogeneous electric fields through the methods of Rydberg-Stark deceleration~\cite{wing80a,breeden81a,hogan16a}. This was first implemented for beams of Kr~\cite{townsend01a} and H$_2$~\cite{yamakita04a}. Subsequent developments~\cite{vliegen04a,vliegen05a} led to the realization of guides~\cite{lancuba13a,allmendinger14a,ko14a,deller16a,deller19a}, velocity selectors~\cite{alonso17a}, lenses~\cite{vliegen06a}, mirrors~\cite{vliegen06b,jones17a}, beam splitters~\cite{palmer17a}, accelerators and decelerators~\cite{vliegen06c,hogan12a,lancuba14a}, and traps~\cite{vliegen07a,hogan08a,hogan09a,hogan12a,lancuba16a}. These have been demonstrated for positronium, H, D, He, Ar, Kr or H$_2$. However, the ubiquity of Rydberg states in almost all atoms and molecules suggests that Rydberg-Stark deceleration can be extended to the preparation of cold samples of heavier and more complex species, provided they can be prepared in hydrogenic Rydberg-Stark states with static electric dipole moments $\gtrsim1000$~D, and lifetimes~$\gtrsim10~\mu$s. 

The methods of Rydberg-Stark deceleration have allowed investigations of excited-state decay processes, including effects of blackbody radiation in H, D, He and H$_2$ on timescales $>1$~ms~\cite{seiler11a,hogan13a,seiler13a,seiler16a,zhelyazkova19a}. They have contributed to developments in positronium physics directed toward precision tests of bound-state quantum electrodynamics and antimatter gravity~\cite{cassidy18a}. And they have enabled studies of the $\mathrm{H}_2^+ + \mathrm{H}_2 \rightarrow \mathrm{H}_3^+ + \mathrm{H}$ reaction at temperatures as low as 300~mK, where contributions from individual angular momentum partial waves were identified~\cite{allmendinger16a}. For atoms in coherent superpositions of Rydberg states the methods of Rydberg-Stark deceleration have permitted the observation of matter-wave interference for particles with dimensions of $\sim300$~nm and electric dipole moments of $\sim11\,000$~D~\cite{palmer19a}.

\begin{figure*}
\begin{center}
\includegraphics[width = 0.98\textwidth, angle = 0, clip=]{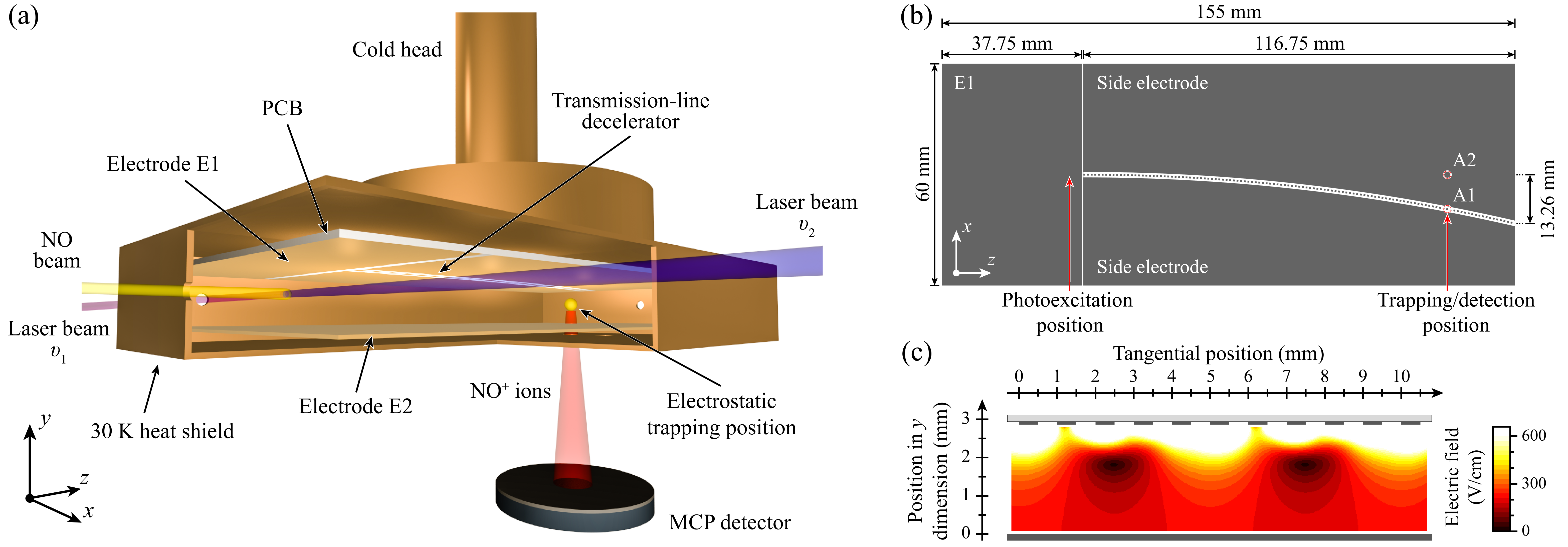}
\caption{(a) Schematic diagram of the apparatus (not to scale). Note: Part of the 30~K heat shield has been omitted for clarity. (b) Scale diagram of the PCB in (a) with silver-coated copper regions (insulating gaps) indicated in grey (white). The positions of detection apertures A1 and A2 in E2 are indicated by the red circles. (c) Typical electric field distribution on the axis of the decelerator in the plane perpendicular to the PCB surface for $V_0=125$~V.}
\label{fig1}
\end{center}
\end{figure*}

Here, by preparing long-lived hydrogenic Rydberg states, for which predissociation is not the dominant decay mechanism, by two-color two-photon excitation from the X\,$^2\Pi_{1/2}$ ground state~\cite{seaver83a,ebata83a,patel07a,deller20a}, we report Rydberg-Stark deceleration and electrostatic trapping of NO. NO is a chemically important radical that plays a role in atmospheric physics, combustion and trace gas detection~\cite{schmidt18a}. Measurements of the decay of the trapped molecules in environments maintained at $T_{\mathrm{env}}=295$ and 30~K, and after initial excitation to states with principal quantum numbers ranging from $n=38$ to $44$, provide new insights into the lifetimes and slow decay processes of Rydberg states of NO.

The apparatus used in the experiments is depicted in Figure.~\ref{fig1}(a). NO molecules in pulsed supersonic beams (mean longitudinal speed $\overline{v}_z=810$~m/s; 25~Hz repetition rate) were photoexcited to high Rydberg states using the resonance-enhanced $\mathrm{X}\,^2\Pi_{1/2}(v''=0, N''=1, J''=3/2)\rightarrow\mathrm{A}\,^2\Sigma^+(v'=0, N'=0, J'=1/2)\rightarrow n\ell\; \mathrm{X}^+\,^1\Sigma^+(v^+=0, N^+)$ two-color two-photon excitation scheme driven by radiation from two pulsed dye lasers (FWHM spectral width 0.17~cm$^{-1}$)~\cite{seaver83a,ebata83a,patel07a,deller20a}. The first, P$_{11}(3/2)$, transition occurred at $\upsilon_1=44\,193.988$~cm$^{-1}$. Transitions from the $\mathrm{A}\,^2\Sigma^+$ state to Rydberg states with $n\ell(N^+)=n$p(0) and $n$f(2) character, where $34\leq n\lesssim60$, occurred for $\upsilon_2=30\,435$ -- 30\,505~cm$^{-1}$. Intramolecular interactions and stray electric fields caused $\sim1\%$ of the excited molecules to evolve, close to the time of excitation, into long-lived $\ell$-mixed hydrogenic Rydberg-Stark states with $|m_{\ell}|\gtrsim3$ and static electric dipole moments between 0 and $\sim10\,000$~D~\cite{deller20a}.

Photoexcitation was performed beneath electrode E1, an electrically isolated 37.75~mm $\times$ 60~mm silver-coated copper region on the printed circuit board (PCB) in Figures~\ref{fig1}(a) and (b). E1 was separated in the $z$ dimension by a 0.5~mm insulating gap from a chip-based transmission-line Rydberg-Stark decelerator~\cite{lancuba14a}. This comprised a curved array of 0.5$\times$0.5~mm silver-coated copper electrodes (1~mm center-to-center spacing, 518.5~mm radius of curvature) on an Arlon substrate. These were separated in the $x$ dimension from side electrodes by 0.5~mm insulating gaps. The PCB was positioned parallel to, and 3~mm above, the plane copper electrode E2 [see Figure~\ref{fig1}(a)]. The decelerator and a surrounding copper heat shield were thermally coupled to a cold head operated at $T_{\mathrm{env}}=295$ or 30~K.

After excitation, the molecules traveled for $5.4~\mu$s ($\sim4$~mm) in the $z$ dimension before the decelerator was activated (rise time $4~\mu$s). At this time those in low-field-seeking (LFS) Rydberg-Stark states with positive Stark shifts were loaded into the first traveling electric trap. The decelerator was operated using five sinusoidally oscillating electric potentials (amplitude $V_0=125$~V; frequency $f_{\mathrm{osc}}=v/L\leq160$~kHz, with $L=5$~mm the spatial periodicity along the decelerator axis) applied to the electrode array in a sequence repeated every sixth electrode, and a potential $-V_0/2$ applied to E2. A typical electric field distribution in the $yz$ plane on the axis of the decelerator is shown in Figure~\ref{fig1}(c). To decelerate molecules confined within the traveling traps, $f_{\mathrm{osc}}$ was reduced toward zero. For $f_{\mathrm{osc}}=0$ the traps remained stationary. The curvature of the decelerator minimized collisions with the trailing components of the molecular beam~\cite{seiler11a}.

Detection was performed \emph{in situ} by pulsed electric field ionization (PFI)~\cite{lancuba16a}. This was implemented by switching off the decelerator and applying potentials of +500~V to the side electrodes. NO$^+$ ions generated above A1, a 2-mm-diameter aperture in E2 at the position of the 101$^{\mathrm{st}}$ decelerator electrode, by PFI were accelerated out from the cryogenic region to a microchannel plate (MCP) detector [see Figure~\ref{fig1}(a)]. To allow detection by PFI when the decelerator was off, a second 2-mm-diameter aperture A2 was included in E2 on the molecular-beam axis [see Figure~\ref{fig1}(b)].

\begin{figure}
\begin{center}
\includegraphics[width = 0.46\textwidth, angle = 0, clip=]{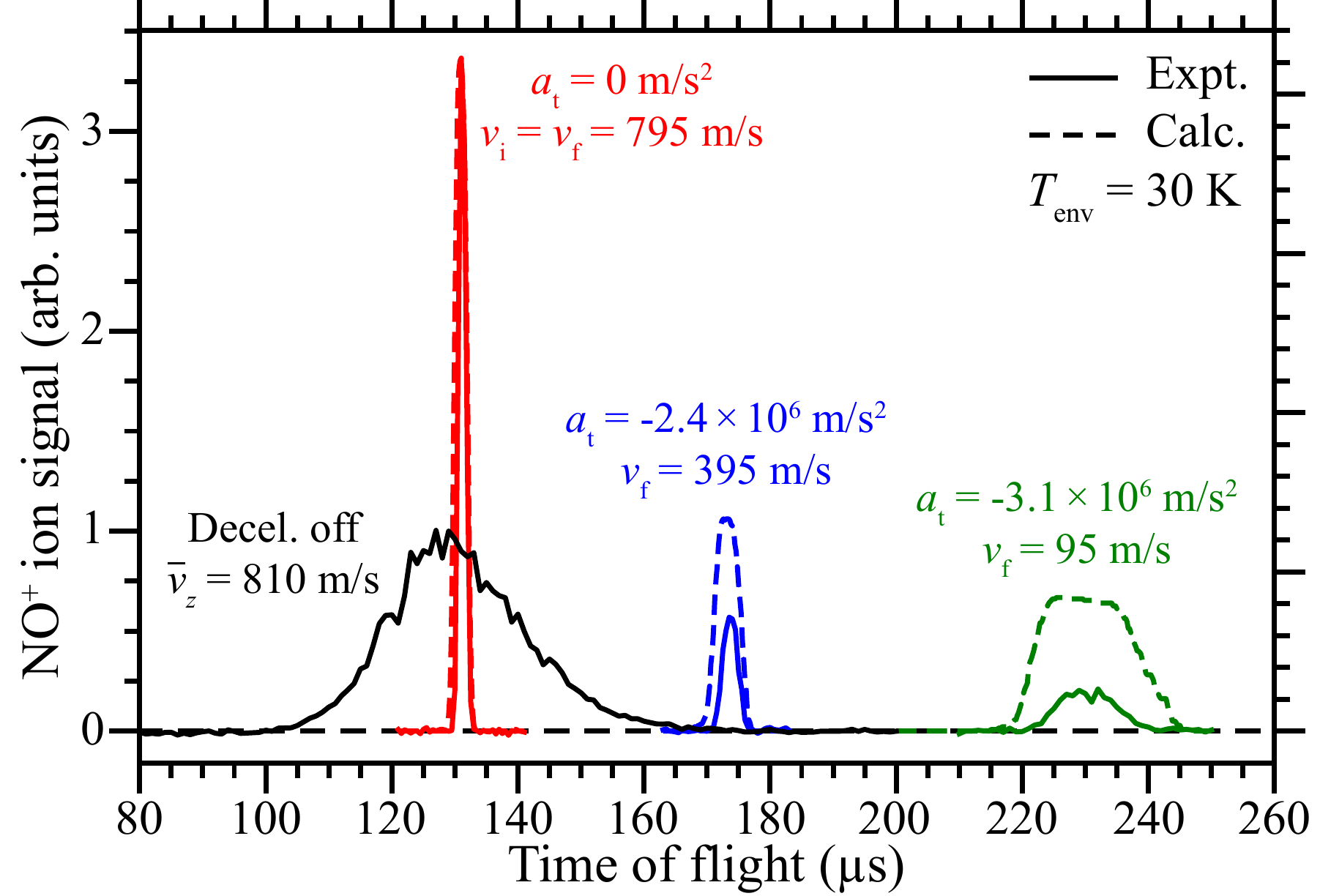}
\caption{Experimentally recorded (continuous curves) and calculated (dashed curves) TOF distributions for Rydberg NO molecules from their position of photoexcitation to the detection position on the molecular beam axis with the decelerator off (black curve), and above aperture A1 for $a_{\mathrm{t}}=0$, $-2.4\times10^6$ and $-3.1\times10^6$ ~m/s$^2$ as indicated.}
\label{fig2}
\end{center}
\end{figure}

To characterize the decelerator it was cooled to $T_{\mathrm{env}}=30$~K and Rydberg states with 43f(2) character were excited. This led to the population of long-lived hydrogenic Rydberg-Stark states with $n=43$ and $N^+=2$ that had static electric dipole moments between 0 and 6400~D. Time-of-flight (TOF) measurements between the photoexcitation and detection positions (105~mm) were then made. A reference measurement performed with the decelerator off is displayed as the black curve in Figure~\ref{fig2}. From the $\sim130$~$\mu$s mean measured flight time, a mean longitudinal speed of $\overline{v}_z=810$~m/s was determined. The decelerator was then activated to transport molecules in LFS Rydberg-Stark states to the detection position above A1 at a constant speed of 795~m/s, i.e., the tangential acceleration was set to $a_{\mathrm{t}}=0$~m/s$^2$. The resulting TOF distribution (red curve in Figure~\ref{fig2}) exhibited a larger amplitude and smaller FWHM than that recorded with the decelerator off. These arose because of the reduced velocity dispersion of the molecules when transported in the traveling traps. 

The molecules were decelerated from 795 to 395~m/s (95~m/s) by chirping the frequency of the oscillating potentials~\cite{lancuba16a} such that $a_{\mathrm{t}}=-2.4\times10^{6}$~m/s$^2$ ($a_{\mathrm{t}}=-3.1\times10^{6}$~m/s$^2$) as seen from the continuous blue (green) curve in Figure~\ref{fig2}. The measured arrival times at the detection position are in good quantitative agreement with the results of numerical calculations of particle trajectories (dashed curves). In these calculations the initial phase-space distribution of 10\,000 molecules was defined from the measured dimensions of the laser beams, geometry of the apparatus, and TOF distributions recorded with the decelerator off. The distribution of Rydberg-Stark states with $n=43$ was described by a Gaussian function centered on the Stark state with zero static electric dipole moment as determined from the electric field ionization measurements reported in Ref.~\cite{deller20a}. The reduction in the amplitude of the TOF signal as $|a_{\mathrm{t}}|$ was increased is a consequence of the reduced phase-space acceptance under these conditions and the geometry of the \emph{in situ} detection region~\cite{lancuba14a}. The discrepancies between the amplitudes of the measured and calculated TOF distributions arise from the finite decay time of the Rydberg states, which was not accounted for in the calculations. The lifetimes of the Rydberg states are influenced by spontaneous emission, intramolecular interactions, transitions induced by blackbody radiation, and the electric fields in the traveling traps. To study these processes in a controlled way it is necessary to bring the traps to rest in the laboratory-fixed frame of reference. 

\begin{figure}
\begin{center}
\includegraphics[width = 0.46\textwidth, angle = 0, clip=]{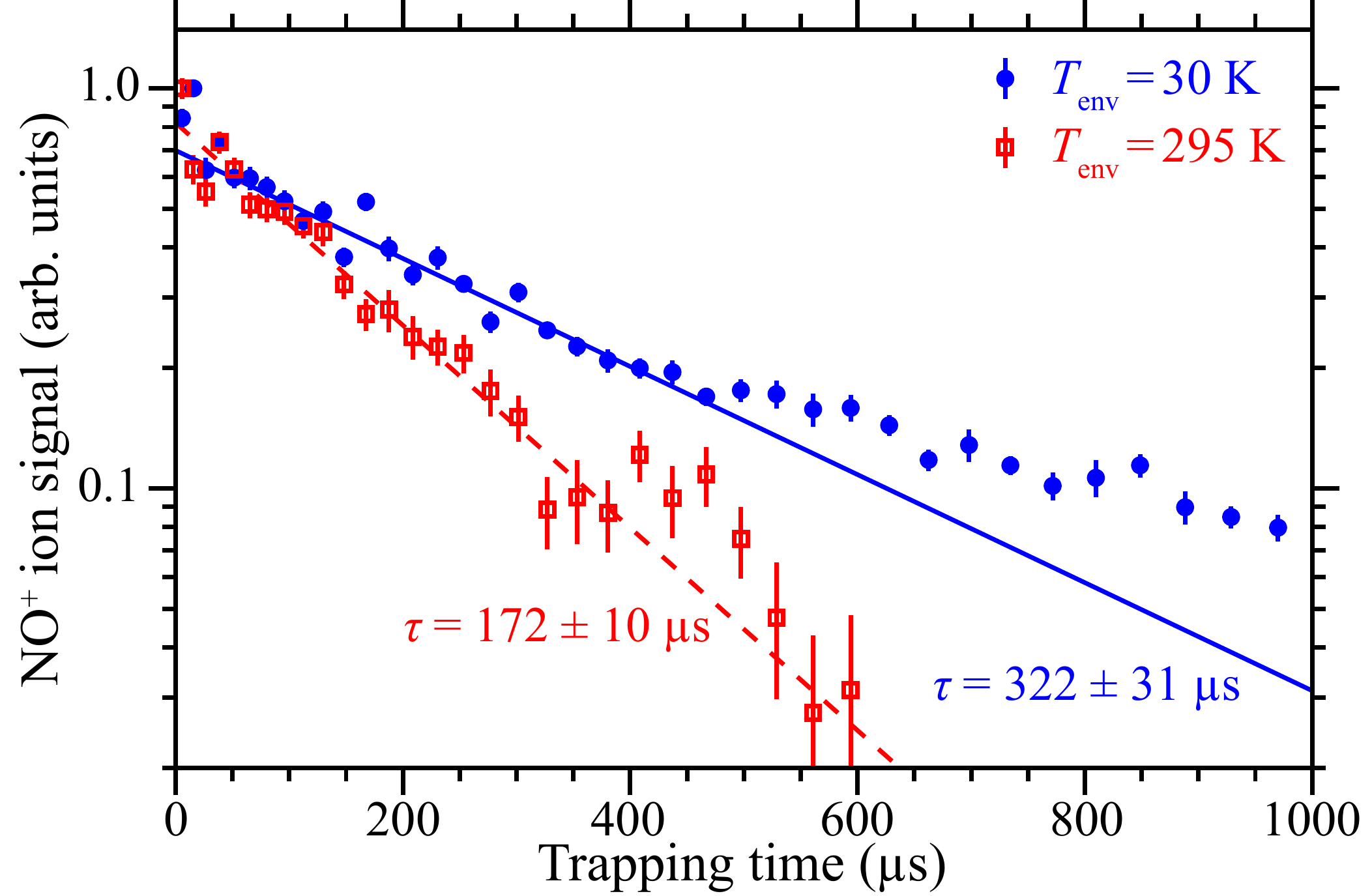}
\caption{Decay of electrostatically trapped Rydberg NO molecules initially excited to states with $n=43$ and $N^+=2$ for $T_{\mathrm{env}}=30$ and 295~K, as indicated. The continuous (dashed) curves represent single exponential functions fit to the data for trapping times between 50 and 350~$\mu$s (see text for details).}
\label{fig3}
\end{center}
\end{figure}

Electrostatic trapping was achieved by setting $a_{\mathrm{t}}=-3.15\times10^{6}$~m/s$^2$ so that the trap in which the molecules were transported was stopped above A1. When stationary, changes in the PFI signal over time reflected motion in, and decay from, the trap. The results of trap-decay measurements performed with molecules prepared in states with $n=43$ and $N^+=2$ are displayed in Figure~\ref{fig3}. Data recorded for $T_{\mathrm{env}}=30$~K (295~K) are indicated by the filled circles (open squares). The origin of the horizontal axis represents the time at which the traps stopped moving. For both measurements, equilibration of the motion of the molecules that occurred after the traps were stopped contributed, in part, to the changes observed in the NO$^+$ signal at trapping times $t_{\mathrm{trap}}\lesssim50~\mu$s.

For $T_{\mathrm{env}}=30$~K, $10-100$ molecules were trapped at a density of $\sim10^4$~cm$^{-3}$ in each experimental cycle. With $V_0=125$~V, the translational temperature of the trapped molecules, measured by monitoring their rate of dispersion after switching off the trapping fields, was $E_{\mathrm{kin}}/k_{\mathrm{B}}=20.5\pm0.8$~K, the trap depth for the outermost $n=43$, $|m_{\ell}|=3$ Rydberg-Stark states was $E_{\mathrm{trap}}/k_{\mathrm{B}}\simeq30$~K, and, as seen in Figure~\ref{fig3} two main decay components were observed for $t_{\mathrm{trap}}\geq50~\mu$s. These are reminiscent of the decay of H Rydberg atoms from cryogenic electrostatic traps~\cite{seiler11a,seiler16a}. The faster component is described by a single exponential function with a time constant of $\tau=322\pm31~\mu$s (continuous blue curve) for $50\leq t_{\mathrm{trap}}\leq350~\mu$s, and attributed primarily to the decay of the $n=43$ and $N^+=2$ Rydberg-Stark states. On longer timescales, transfer to high-$|m_{\ell}|$ states through collisions or interactions with time-dependent electric fields after photoexcitation~\cite{seiler16a}, and effects of blackbody-radiation-induced transitions become apparent. Blackbody transitions have an approximately equal probability of increasing or decreasing the values of $n$ and $|m_{\ell}|$, but preserve the electric dipole moments and LFS character of the molecules, allowing them to remain trapped. However, because lower $|m_{\ell}|$ states decay more rapidly, at longer trapping times only molecules in longer-lived higher $|m_{\ell}|$ states remain. Consequently, the rate of decay from the trap reduces. For $T_{\mathrm{env}}=295$~K the number of trapped molecules was lower than for $T_{\mathrm{env}}=30$~K. To allow direct comparison, the data in Figure~\ref{fig3} have been normalized. Those recorded for $T_{\mathrm{env}}=295$~K were truncated at 0.6~ms where the signal had decayed to the level of the background noise. For $50\leq t_{\mathrm{trap}}\leq350~\mu$s the data recorded for $T_{\mathrm{env}}=295$~K can be described by a single exponential function with $\tau=172\pm10~\mu$s (dashed red curve). Under these conditions, the decay of the molecules from the trap is affected by blackbody photoionization~\cite{seiler11a,deller19a}, and slow dissociation following blackbody transitions to states with $|m_{\ell}|\lesssim3$. 

\begin{figure}
\begin{center}
\includegraphics[width = 0.49\textwidth, angle = 0, clip=]{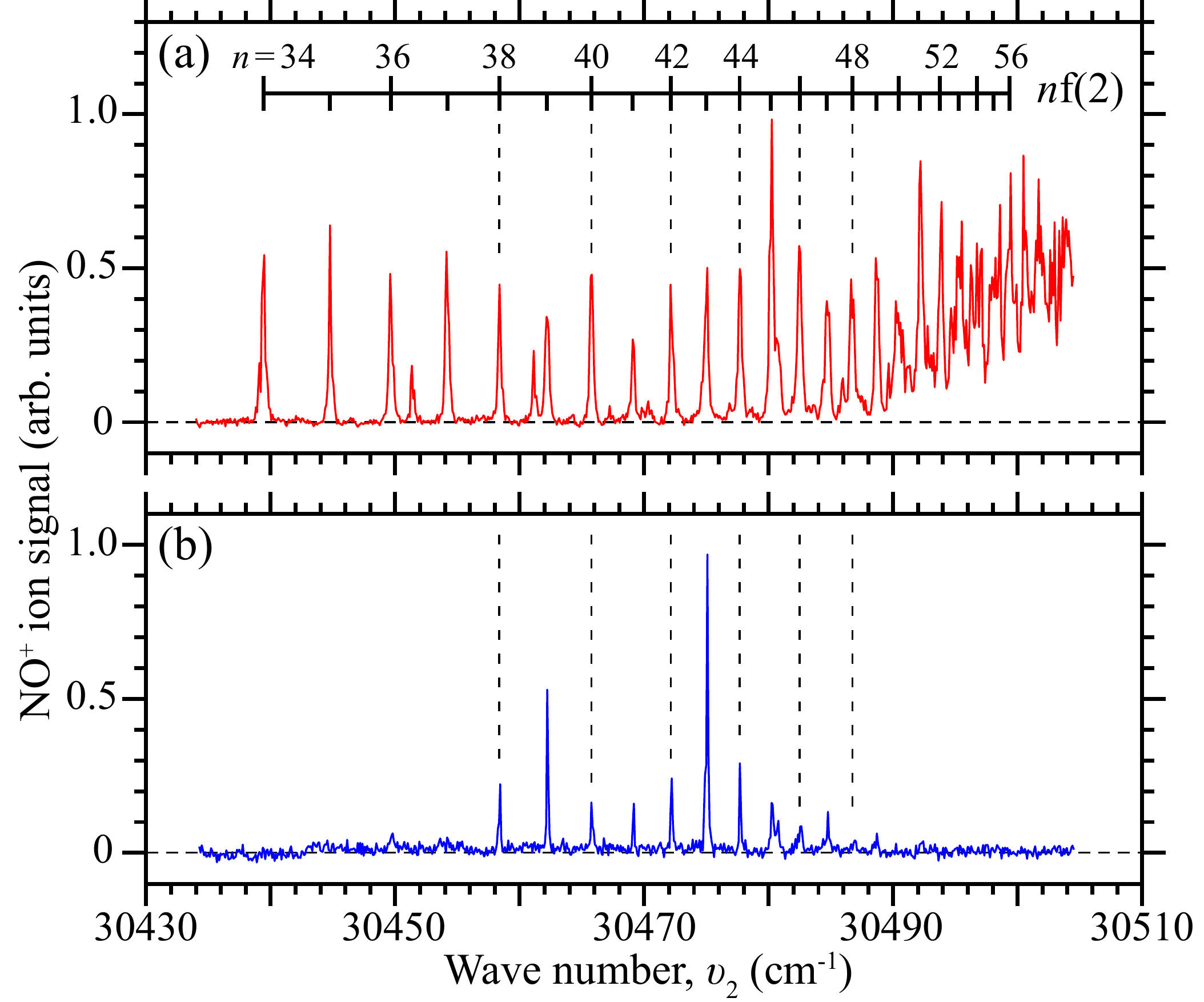}
\caption{Laser photoexcitation spectra recorded (a) upon detection by PFI $\sim50$~ns after photoexcitation, and (b) after deceleration and trapping for $20~\mu$s with $V_0=125$~V and $T_{\mathrm{env}}=30$~K.}
\label{fig4}
\end{center}
\end{figure}

For $V_0=125$~V, molecules in states with values of $n$ from 38 to 49 could be decelerated and trapped. This Rydberg-state acceptance was determined from the spectra in Figure~\ref{fig4}. The spectrum in Figure~\ref{fig4}(a) was recorded by PFI $\sim50$~ns after photoexcitation and is dominated by transitions to $n$f(2) Rydberg states. This is because, for the values of $n$ studied, the short predissociation times of the $n$p(0) states ($<1$~ns~\cite{vrakking95a}) preclude their detection. The spectrum in Figure~\ref{fig4}(b) was obtained by detecting only molecules that were decelerated and trapped for $20~\mu$s at $T_{\mathrm{env}}=30$~K. For the lowest (highest) values of $n$ the trapping efficiency reduces in part because the maximal electric dipole moments (ionization electric fields) reduce and greater losses occur during deceleration. The relative intensities of the individual spectral features reflect the efficiencies with which long-lived hydrogenic Rydberg-Stark states were populated upon excitation at each resonance, and the contributions from intramolecular interactions that cause slow dissociation during deceleration and trap loading.

\begin{figure}
\begin{center}
\includegraphics[width = 0.41\textwidth, angle = 0, clip=]{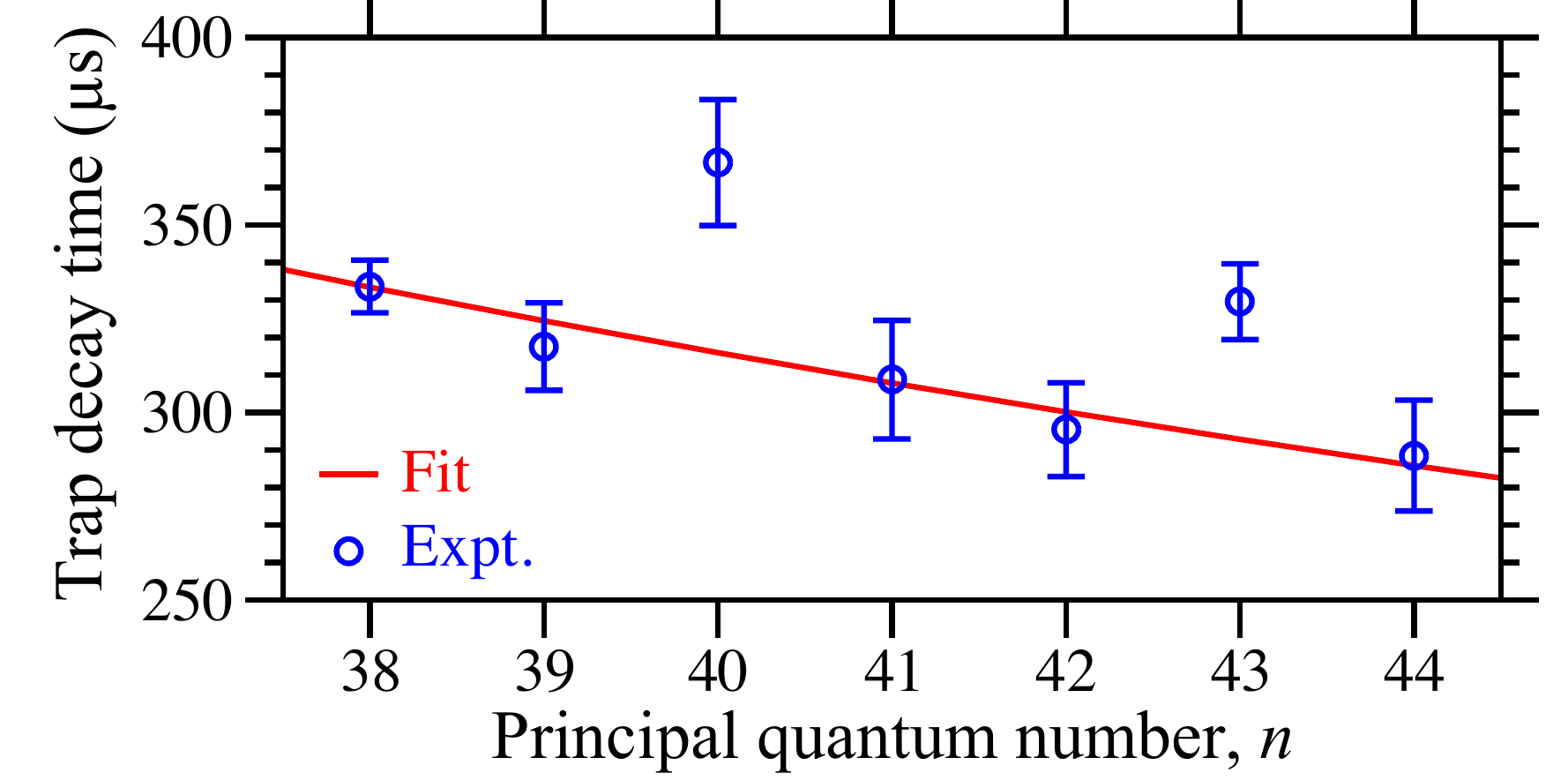}
\caption{Rydberg-state dependence of $\tau$ determined for states with $N^+=2$ and $50\leq t_{\mathrm{trap}}\leq 350~\mu$s with $V_0=125$ and $T_{\mathrm{env}}=30$~K (open blue circles). The continuous red curve for which $\tau=15197\,n^{-1.05}~\mu$s was fit to the experimental data for values of $n$ excluding 40 and 43 (see text for details).}
\label{fig5}
\end{center}
\end{figure}

The decay of molecules with $38\leq n\leq44$ and $N^+=2$ was measured for $T_{\mathrm{env}}=30$~K, and $\tau$ was determined for $50\leq t_{\mathrm{trap}}\leq350~\mu$s. These decay times (blue points in Figure~\ref{fig5} that represent the average of multiple measurements) are shorter than the corresponding fluorescence lifetimes, $\tau_{\mathrm{fl}}$, of $\ell$-mixed Rydberg-Stark states with $|m_{\ell}|=3$ in the H atom for which $\tau_{\mathrm{fl}}\propto n^4$ and $\tau_{\mathrm{fl}}(n=38)=547~\mu$s. In general, the decay times in Figure~\ref{fig5} reduce as the value of $n$ increases. The red curve represents the function $15\,197\,n^{-1.05}~\mu$s fit to the data for values of $n$ excluding 40 and 43. This behavior is not representative of the trap decay observed for atomic Rydberg states and is attributed to the gradual increase in the strength of intramolecular interactions of the high-$n$ hydrogenic Rydberg-Stark states in the series with $N^+=2$ and $v^+=0$, with short-lived states in series converging to other rotational or vibrational states of the NO$^+$ ion core as the $N^+=2$ series limit is approached.

The $n=40$ and 43 states in Figure~\ref{fig5} exhibit trap decay times that do not follow this $n^{-1}$ scaling. These states are almost degenerate with the hydrogenic Rydberg-Stark states in the $N^+=0$ series for which $n=44$ and 48, respectively. Mixing between these near degenerate Stark manifolds, for which $\Delta N^+=2$, arises as a result of the charge-quadrupole interaction of the Rydberg electron with the NO$^+$ ion core~\cite{bixon96a}. The enhanced trapping times for these values of $n$ appear to be a consequence of these intramolecular interactions and indicate that the $n=44$ and $48$ states in the $N^+=0$ series exhibit longer lifetimes than the $n=40$ and $43$ states with $N^+=2$.

In conclusion, we have demonstrated electrostatic trapping of NO molecules in long-lived hydrogenic Rydberg-Stark states. Significant differences in the trap decay times for $T_{\mathrm{env}}=295$ and 30~K highlight effects of blackbody photoionization, and slow dissociation following blackbody induced transitions between Rydberg states. Measurements performed for a range of values of $n$ yielded new insights into effects of intramolecular interactions on slow decay process of long-lived Rydberg states in NO. 

\begin{acknowledgments}
This work was supported by the European Research Council (ERC) under the European Union's Horizon 2020 research and innovation program (Grant No. 683341). M.\,H.\,R is grateful to the Engineering and Physical Sciences Research Council for a Vacation Bursary, and support through the Doctoral Training Partnership (Grant No. EP/R513143/1).
\end{acknowledgments}
	

\end{document}